\documentclass[preprint,showpacs,preprintnumbers,jap]{revtex4}
\usepackage{amsfonts}
\usepackage{amsmath}
\usepackage{graphicx}
\usepackage{dcolumn}
\usepackage{bm}

\begin{document}

\title{AC voltage effects on indirect excitons in a coupled quantum-well pair}
\author{H. Cruz \\
{\it Departamento de F\'\i sica, }\\
{\it Universidad de La Laguna, 38204 La Laguna, Tenerife, Spain.}}

\begin{abstract}
The objective of this work is to numerically integrate in space and time the effective-mass Schr\"odinger equation for an excitonic wave packet in a coupled quantum-well system under a time-dependent electric field. Taking as a starting point a time-dependent Hartree potential, we derive the nonlinear dynamical evolution of the excitonic wave function. 
We found that the system resonant condition can be modified owing to the reaction field generated by a charge dynamically trapped in the quantum-well pair.
As a result, this study raises a broader question: is it possible to create an electrostatic trap for indirect excitons at high ac frequencies? 
We conclude there is a possibility of having another kind of terahertz electromagnetic radiation emerging from an excitonic trap.

{\bf PACS number(s): }73.20.Dx Electron states in low-dimensional
structures, 73.40.Gk Tunneling, 73.50.-h Electronic transport phenomena in
thin films.
\end{abstract}
\maketitle

\section{Introduction}

Recently, there has been increasing interest in the study of the
optical and electrical properties of semiconductor nanostructures in semiconductor physics [1--3]. 
Many investigators have focused on the properties of indirect excitons in semiconductor quantum wells.
Indirect excitons in a coupled quantum-well pair are bound hole and electron pairs. 
The pairs consist of 
a hole and a electron spatially separated into two different quantum wells [4]. 
As a result, the dipolar excitons form
a bosonic system of interacting quasiparticles.
However, efficient trapping of high
excitons densities [5--6] is required to obtain a Bose-Einstein condensation of excitons in the 2D plane of a nanostructure.
Such exciton traps have been achieved
within last years. These include disorder traps [7], strain-induced traps [8--10],
magnetic traps [11], laser induced traps [12] and acoustic traps [13].

Nevertheless, one of the the most promising exciton traps is the quantum-confined Stark effect in
semiconductor quantum wells [14--22].
In this case, 
a quantum-well pair is used to obtain the active region in the device. 
With external laser radiation, it is possible to create electron-hole pairs 
in the semiconductor layers. Then, the pairs can
relax via optical and acoustic phonon emission to form spatially indirect excitons. The hole and electron are 
spatially separated in different semiconductor layers.
Many studies have highlighted that 
the energy relaxation time to the ground state is significantly shorter than the recombination lifetime of the indirect excitons. 
These excitons can cool down to lattice temperatures before recombining. 
Thus, the indirect excitons exhibit lifetimes of up to microseconds, which are determined by the spatial overlap of the electron
and hole wave function. By contrast, the direct
excitons recombine on nanosecond time scales.

In experiments (Fig. 1.b), 
an indirect exciton is a bound state of an electron and a hole in different semiconductor layers.
The carriers are separated by 
a distance of nanometers under an external electric field ($F$ is the module
of the external electric field).
At $F=0$, we have a symmetric double well (Fig. 1.a) and the electrons and holes can tunnel
through the central barrier in the trap. So, we have direct excitons that recombine on a nanosecond time scale at $F=0$
in a coupled quantum-well pair.
As a result, a tunable trap can be created using the easy voltage tunability of the quantum-confined Stark effect 
for indirect excitons (Fig. 1.b). 
The tunable voltage can be applied through external electrodes.

As can be seen in Ref. [23], there is
the possibility of having an
electrostatic trap for indirect excitons at $F$=0.
At high $n_s$ values ($n_s=3\times 10^{11}$cm$^{-2}$),
they demonstrated 
electrostatic trapping of indirect excitons in the absence of an external
electric field.
In such an exciton trap, the symmetry of the double quantum-well potential was broken because of charge build-up
effects at $F$=0. An electrostatic trap was then found to be possible for indirect excitons in the absence of an
external electric field. 
The internal charge distribution in the device modified the potential across the semiconductor layers.
The tunneling oscillations between the two quantum wells vanished because of the Hartree potential difference between the two quantum wells.
In addition to this, Winbow {\it et. al.} studied ac voltage effects on indirect excitons [24]. They reported on indirect excitons in conveyors created by a set of ac voltages. 
They were able to achieve a moving lattice of indirect excitons.
The wavelength of this moving lattice was set by the
electrode periodicity, the amplitude was controlled by the applied voltage
and the velocity was controlled by the ac frequency. They founded the dynamical
localization transition for excitons in the conveyors and determined its
dependence on exciton density.

It was noted earlier that one remaining key question in these experiments is 
the theoretical analysis of ac field effects on the dynamical processes.
The ac field can be applied between the external electrodes of the device.
We know that 
if an external ac field is applied perpendicular to the plane of the
semiconductor layers,
electrical properties can be strongly modified because of voltage-buildup effects. 

Taking into account this information, extending the analysis to a coupled quantum-well pair considering 
carrier-carrier interactions appears interesting.
With this is mind, we studied the time-dependent 
evolution of the electron and hole wave packets
in a coupled quantum-well pair under external ac fields. The
method of calculation was based on the discretization of space and time
for both electron and hole wave functions. We show here that the system resonant condition 
can be strongly modified by ac voltage effects. 
We shall see below that there is a possibility of having an electrostatic trap for indirect
excitons at high ac frequencies. 

This paper is organized as follows. Section II describes the model used in this work. In section III the
results are discussed.

\section{Model}

In order to study the charge density dynamics in the structure growth
direction, we need to solve the time-dependent Schr\"odinger equation
associated with an electron-hole pair in an exciton trap under an external ac field [25]. 
The exciton trap will be formed by a coupled quantum-well pair.
Assuming translation symmetry of the Hamiltonian in the $xy$ plane perpendicular to the growth direction, 
the electron $\psi _{e}$ and hole $\psi _{h}$ wave functions in
the $z$ axis will be given by the nonlinear Schr\"{o}dinger equations
[26] 
\begin{equation}
i\hbar \frac{\partial }{\partial t}\psi _{e}(z,t)=\left[ -\frac{\hbar ^{2}}{%
2m_{e}^{\ast }}\frac{\partial ^{2}}{\partial z^{2}}+V_{e}(z)+V_{H}\left(
\mid \psi _{e}\mid ^{2},\mid \psi _{h}\mid ^{2}\right) 
+eFz sin \left( \frac{2 \pi t}{T} \right) \right] \psi
_{e}(z,t),  \label{electron}
\end{equation}

\bigskip\ 
\begin{equation}
i\hbar \frac{\partial }{\partial t}\psi _{h}(z,t)=\left[ -\frac{\hbar ^{2}}{%
2m_{h}^{\ast }}\frac{\partial ^{2}}{\partial z^{2}}+V_{h}(z)-V_{H}\left(
\mid \psi _{e}\mid ^{2},\mid \psi _{h}\mid ^{2}\right) 
- eFz sin \left( \frac{2 \pi t}{T} \right) \right] \psi
_{h}(z,t),  \label{hueco}
\end{equation}%
where 
$F$ is the external electric field. The ac field is applied perpendicular to 
the semiconductor layers and $T$ is 
the ac oscillation period.  
The subscripts $e,h$ refer to electrons or holes, respectively, and $%
V_{e}^{{}}(z)$ and $V_{h}^{{}}(z)$ are the potentials established by the coupled quantum-well pair.
The $m_{e}^{\ast }$ and $m_{h}^{\ast }$ are the electron and hole effective
masses, respectively, and $V_{H}$ is a Hartree potential given by the
electron-hole interaction in the quantum-well region. 
It is interesting to note that the Hartree term is a nonlinear potential that
depends on the wave function form.
The Hartree
potential is given by the Poisson's equation [26]
\begin{equation}
\frac{\partial ^{2}}{\partial z^{2}}V_{H}(z,t)=-\frac{e^{2}n_{s}}{%
\varepsilon }\left[ \left\vert \psi _{e}(z,t)\right\vert ^{2}-\left\vert
\psi _{h}(z,t)\right\vert ^{2}\right] ,  \label{poisson}
\end{equation}%
where $n_{s}$ is the
carrier sheet density and $\varepsilon $ is the permittivity with respect to the vacuum. 
We should point out that the model underlying this work is based on single electron-hole pairing
between the conduction and valence bands.
In addition to this, we can note that 
there are many-body carrier interactions in the $xy$ plane, i.e., the interaction between two different electron-hole pairs. 
In that case, two electron-hole pairs localized in different positions can interact between each other
through a dipole-dipole force.
By contrast,
we know that many-body interactions are much smaller than the single electron-hole Coulomb term
for our $n_{s}$ values.
In order to simplify our study, only a single electron-hole pairing term will be considered in this article.
The dynamics in the $xy$ plane will not be considered in this study.
Moreover, it is clear that Hartree theory neglects contributions to the energy beyond the exchange term and is therefore expected to overestimate the carrier-carrier potential value. 
The local-density approximation of the functional theory is teh simplest approach to consider energy contributions
beyond the exchange term.
These contributions will be neglected in our model.

In this work, we consider a symmetric double quantum well
potential where the barriers
between the wells and the electrodes consist of Ga$_{1-x}$Al$_{x}$As and the left and right wells consist of GaAs.
In absence of external field, we have a symmetric double quantum well (Fig. 1.a), where the carrier wave functions
are extended along the quantum wells.
It is interesting to note that a 
tunneling condition between
both symmetric wells is obtained at $F=0$ [27]. Consequently, both electron and
hole quantum states are aligned, so tunneling oscillations between both wells are allowed.
In the presence of an external electric field (Fig. 1.b), carrier states in the bands
are predominantly localized in their respective wells. 
The coupling between the wells is very weak and the carrier oscillations are strongly reduced.
Moreover, the quantum-well pair enables optical transitions in
both wells to be discriminated by their different spectral positions [27] if an electric field is applied (Fig. 1.b).
In Fig. 1.b,, electrons are photoexcited in the right well, while holes are localized in the left layer.

The Eqs. (1), (2) and (3) are solved by applying the split-step method [28].
To model the nonlinear dynamics we apply simplifying assumptions and standard approximations.
In the split-step approach, the wave packet is
advanced in time steps $\delta t$. The steps are short enough that the algorithm $%
e^{-i\delta tT_H/2}e^{-i\delta tU}e^{-i\delta tT_H/2}$ can be applied to the
generator. 
The kinetic and potential energies are segregated on separated generators. 
$T_H$\ and $U$ are the kinetic and potential terms of the Hamiltonian
and $\delta t$ is the time step. 
The algorithm freely propagates the wave packet for $\delta t/2$, applies the full potential interaction, then freely propagates the wave packet for the remaining $\delta t/2$
in each time step $\delta t$. 
The method of calculation
is stable and norm preserving.
It has been widely employed to study
time-dependent
Hamiltonian problems [29]. 
The photoexcitation of an electron wave packet in the right well creates a hole wave
packet in the left well at the same time (Fig. 1.a), which will affect the behavior of the
electron dynamics through the Coulomb interaction.
With this in mind, we also solved
the Poisson equation associated with $V_H$ using another tridiagonal
numerical method for each ${\it t}$ value [28].

\section{Results and discussion}

We numerically integrated Eqs. (1) and (2) using an
initial 2D carrier sheet density equal to $n_s=3\times 10^{11}$ cm$^{-2}$, $T=$ 10 ns
and $F=20$ kV/cm.
The initial probability has been taken to be 1.
In Figs. 2 and 3, 
we have plotted 
the amplitude of both electron and hole wave
functions $\left| \psi _{e,h}\right| ^2$ and both conduction and valence
band potentials. 
According to previous works, both electron and hole wave functions are initially created in the center of the left and right quantum wells, respectively, at $t=0$. 
We considered initial wave packets with zero average momentum. The quantum dynamics is evaluated by solving Eqs. (1) and (2).
We based the calculations on a GaAs/Ga$_{1-x}$Al$%
_{x}$As double quantum-well system which consists of two 80 {\AA } wide GaAs quantum wells separated by a barrier of thickness 20 {\AA }. 
In all numerical examples, $m_{e}$=0.067$m_{0}$ and $m_{h}$=0.7$m_{0}$ where $m_{e}$ and $m_{h}$ are the
electron and hole effective-masses, respectively.
The equations are then numerically solved using a spatial mesh size of 0.5
\AA ,\ a time mesh size of 1.0 a.u., and a finite box (1350 {\AA }) large
enough so as to neglect border effects. 
Both electron and hole wave functions are seen to spread out along the semiconductor region as time passes. Owing to the lower effective mass, the electron wave function spreads out more rapidly. 

In view of the above, 
the numerical integration in time allows us to obtain the 
probability density of finding carriers, $P_{a,b}$, in a device region [$a$, $b$]
at any time $t$ 
\begin{equation}
P_{a,b}(t)=\int_a^bdz\ |\psi_{e,h} (z,t)|^2,
\end{equation}
where [$a$,$b$] are the quantum well limits. Figs. 4 and 5 show the time evolution of the electron and hole probability densities in the left and right quantum wells, respectively. 
In Fig. 4, we took a low charge density value, 
$%
n_s=0.15\times 10^{11}$ cm$^{-2}$ and $F=$0 kV/cm. The $n_s$ values were obtained through Eq. (4). 
As depicted in Fig. 4, it is found 
the existence of tunneling
oscillations between both quantum wells. 
At $%
n_s=0.15\times 10^{11}$ cm$^{-2}$, 
the amplitude of the oscillating charge density is found to be approximately equal
to 1.

As can be seen in Fig. 5, the amplitude of the oscillations is strongly reduced at higher density values.
Such a result can be easily explained within the framework of the model.
It is clear that the carrier energy levels of both wells are
exactly aligned at $n_s=0 $ cm$^{-2}$ in a double quantum-well system (Fig. 1.a). 
Particularly, the total charge density will oscillate between both wells with a certain
tunneling period (Fig. 4). 
However, both eigenvalues are not aligned if the
carrier density takes a higher value 
$n_s=3\times 10^{11}$ cm$^{-2}$  
(Fig. 5). In that case, the amplitude of the oscillating charge is strongly decreased. 
The tunneling oscillations between the two wells vanish owing to the Hartree potential difference between the two quantum wells.
The symmetry of the coupled quantum-well pair is broken due to charge build-up
effects. 
An electrostatic trap is then found to be possible for indirect excitons at $F=0$. 

Figs. 6 and 7 show carrier oscillations versus time at $T=10$ ns and at $T=50$ ns,
respectively. $T$ is the ac oscillation period and $F$ is the module of
the electric field, i.e., $F=20$ kV/cm.
As can be seen in Figs. 5 and 6, both density curves are identical if the oscillation
period is small enough. The applied electric field produces no significant change in
the dynamics at $T=$10 ns. As a result, electrostatic trapping of indirect excitons is
obtained at high ac frequencies.

In Fig. 8, we plotted the amplitude of the electron tunneling oscillations versus $T$. 
The amplitude value reached a peak at $T=50$ ns.
If $T<50$ ns, we found that the amplitude of the tunneling oscillations increases as we raise
$T$, up to a resonant value of the oscillation period ($T_r =50$ ns). 
Figs. 7 and 9 give the carrier oscillations versus time at resonant 
oscillation period ($T_r= 50$ ns)
and at $T=90$ ns, respectively.

As can be seen in Fig. 7, we have two new effects: one, 
the existence of a resonant tunneling peak at a resonant $T_r$ value,
and the other, the decline of the amplitude of the oscillating
charge below 1 at $T=T_r$ (Fig. 7).
Let us study the first. 
At low $T$ values, the $F$ oscillations are so fast that the wave packets
do not have time to tunnel through the barrier. 
However, the amplitude of the
electron oscillations increases as we raise $T$, up to $T=T_r$ (Fig. 7). We note that the hole
wave function cannot tunnel through the barrier at resonant condition
because of the higher hole effective-mass.
At $T>T_r$ values, the existence of tunneling in the valence band can be
seen, as shown by Fig. 9. Then, the time-dependent evolution of the electron wave
packet is modified because of the $V_H$ potential in the valence band, i.e.,
because of nonlinear effects.

Let us now discuss the second effect. We need to consider our nonlinear
effective-mass Schr\"odinger equation to explain this result.
The time-dependent evolution of the electron wave
packet is strongly modified because of the
attractive Hartree potential in Eq. (3) at $n_s>0$. 
We know that the amplitude of the tunneling oscillations is equal to
approximately 1 at the resonant condition in the absence of carrier-carrier
interaction [28].
However, 
current research has confirmed that
this result can be modified because of nonlinear effects.
A reaction field that modifies the amplitude of
the tunneling oscillations is produced by the 
charge dynamically trapped in the
coupled quantum-well pair. 
As can be seen in Ref. [28], the amplitude of the
tunneling oscillations is never equal to 1. 

Fig. 10 gives the carrier oscillations versus time at $n_s=3\times 10^{11}$ cm$^{-2}$, $F=$20 kV/cm 
and $T=$120 ns. 
It is interesting to note that
electron tunneling oscillations
are not uniform in Fig. 10. The electron oscillations in the conduction
band are strongly influenced by the ac frequency. 
Clearly, the external electric field 
modulates the electron oscillations. 
We have coherent tunneling between both quantum wells because of 
the superposition of both symmetric and antisymmetric
quantum-well eigenstates.
As a result, a time-varying excitonic dipole moment that allows the
emission of electromagnetic radiation at the oscillation frequency
is obtained. 
In Fig. 10, we have a charge
density that oscillates with two different periods and thus,
we have electromagnetic radiation at two different oscillation
frequencies. 
This result leads us to conclude there is a possibility of having 
terahertz radiation emerging
from the sample with two different frequencies. 
Owing to
their possible applications as high-frequency emitters,
considerable
attention to these structures can be expected.

In Fig. 11, we plotted the hole probability density versus time at $T=$150 ns.
As can be seen in Fig. 11, the
existence of tunneling oscillations in the valence band is found.
Such a result can be easily explained as follows:
the electric field oscillations are so slow that the hole wave packet has time 
to tunnel through the barrier at $T=$150 ns.
We note that hole tunneling time takes a higher value because of 
the higher hole effective-mass (Fig. 4).
As a consequence, the amplitude of the hole oscillations is strongly increased at low
ac frequencies. If both wave packets can tunnel through the barrier, we have 
direct excitons that recombine on a nanosecond time scale.

Due to the fact that the
the mean life for an electron-hole pair in GaAs is
of the order of several hundreds of picoseconds, in principle, an experimental
observation of such a process is possible.
The recombination time of the direct excitons is significantly shorter than
the lifetime of the indirect excitons.
In principle, an electric field can be applied to create a tunable trap
(Fig. 1.b). 
With laser radiation, it is 
possible to create electron-hole
pairs that can relax to form spatially indirect excitons.
Next, the time-independent electric field can be switched off. If the carrier density
is large enough, the tunneling oscillations between the quantum wells
will vanish due to the carrier-carrier interaction in the semiconductor region.
The symmetry of the double quantum well is broken owing 
to charge build up effects.
An ac electric field can now be applied to the quantum wells.
Similarly, the lifetime time of excitons could be determined
experimentally when the ac electric field is switched on.
At high $T$ values, the recombination time for an electron-hole pair will be of the
order of the lifetime for direct recombination.
Alternatively, if the ac oscillation period is small enough, the recombination lifetime
will be significantly longer.

Finally, in this work we numerically integrated in space and time a
nonlinear effective mass Schr\"odinger equation in a coupled quantum-well
pair under an external ac field. In the same way, interaction between carriers were considered in our model
through a time-dependent Hartree potential. 
It can be seen that
the nonlinear
carrier dynamics are determined by three different competing potentials: 
the electron-hole potential, the ac voltage potential and the quantum-well potential.
If the ac oscillation period is small enough, we showed the possibility of an electrostatic trap 
forming for indirect excitons.
The applied field produces no significant change in the dynamics at high ac frequencies.
Moreover, it is found that the amplitude of the tunneling oscillations in the conduction band increases as
we raise $T$, up to a resonant value of the ac frequency. At low ac frequencies, we have
direct excitons in a coupled quantum-well pair that recombine on a nanosecond time scale.
In addition, we obtained two time-varying
moments in the heterostructure with two different frequencies at certain $T$ values.
We conclude there is a the possibility of having another kind of terahertz photonic radiation
emerging from a quantum-well device.

\newpage

\section{Figures}

\begin{itemize}
\item {\bf Fig. 1} (a) Energy band diagram of the coupled quantum well in absence of external 
electric field. $F$ is the module of the external electric field.
(b) An indirect exciton in coupled quantum wells is a bound state of 
an electron an a hole in separate wells.

\item {\bf Fig. 2} Conduction and valence band potentials and carrier wave
functions at $t=3$ ps and $F=20$ kV/cm. We took an initial 2D carrier sheet density
equal to $n_s=3\times 10^{11}$ cm$^{-2}$ and $T=$10 ns.
We considered a GaAs/Ga$_{1-x}$Al$%
_{x}$As double quantum-well system which consists of two 80
{\AA } wide GaAs quantum wells separated by a barrier with thickness of 20 \AA. 

\item {\bf Fig. 3} Conduction and valence band potentials and carrier wave
functions at $t=4$ ps and $F=20$ kV/cm. We took an initial 2D carrier sheet density
equal to $n_s=3\times 10^{11}$ cm$^{-2}$ and $T=$10 ns. We 
considered the double quantum-well system described in the caption of Fig. 2.

\item {\bf Fig. 4} Probability density in the left (electron) and right (hole) quantum wells ($P_{ab}$) versus time. We took an initial 2D electron sheet density equal to $%
n_s=0.15\times 10^{11}$ cm$^{-2}$ and $F=$0 kV/cm. Thin line: electron. Thick line: hole. We 
considered the double quantum-well system described in the caption of Fig. 2.

\item {\bf Fig. 5} Probability density in the left and right quantum wells ($P_{ab}$)
versus time. We took an initial 2D electron sheet density equal to $%
n_s=3\times 10^{11}$ cm$^{-2}$ and $F=$0 kV/cm. We considered the double quantum-well system described in the caption of Fig. 2. 

\item {\bf Fig. 6} Probability density in the left and right quantum wells ($P_{ab}$)
versus time. We took an initial 2D electron sheet density equal to $%
n_s=3\times 10^{11}$ cm$^{-2}$, $F=$20 kV/cm and $T=$10 ns. We considered the double quantum-well system described in the caption of Fig. 2.

\item {\bf Fig. 7} Probability density in the left and right quantum wells ($P_{ab}$)
versus time. We took an initial 2D electron sheet density equal to $%
n_s=3\times 10^{11}$ cm$^{-2}$, $F=$20 kV/cm and $T=$50 ns. We considered the double quantum-well system described in the caption of Fig. 2.  

\item {\bf Fig. 8} Amplitude of the tunneling oscillations versus $T$ in
the conduction band. We considered the double quantum-well system described in the caption of Fig. 2.

\item {\bf Fig. 9} Probability density in the left and right quantum wells ($P_{ab}$)
versus time. We took an initial 2D electron sheet density equal to $%
n_s=3\times 10^{11}$ cm$^{-2}$, $F=$20 kV/cm and $T=$90 ns. We considered the double quantum-well system described in the caption of Fig. 2. 

\item {\bf Fig. 10} Probability density in the left and right quantum wells ($P_{ab}$)
versus time. We took an initial 2D electron sheet density equal to $%
n_s=3\times 10^{11}$ cm$^{-2}$, $F=$20 kV/cm and $T=$120 ns. We considered the double quantum-well system described in the caption of Fig. 2. 

\item {\bf Fig. 11} Probability density in the left and right quantum wells ($P_{ab}$)
versus time. We took an initial 2D electron sheet density equal to $%
n_s=3\times 10^{11}$ cm$^{-2}$, $F=$20 kV/cm and $T=$150 ns. We considered the double quantum-well system described in the caption of Fig. 2.

\end{itemize}


\newpage

\begin{thebibliography}{9}

\bibitem{01} W. Hong-wei, M. Xian-wu, H. Yong-gang and S. Ke-hui,
J. Appl. Phys. {\bf 113}, 043105 (2013).

\bibitem{02} D. Churochkin, R. McIntosh and S. Bhattacharyya,
J. Appl. Phys. {\bf 113}, 044305 (2013).

\bibitem{03} P. Manousiadis, S. Gardelis, and N. G. Nassiopoulou
J. Appl. Phys. {\bf 113}, 043703 (2013).

\bibitem{1} G. J. Schinner, E. Schubert, M. P. Stallhofer, J. P. Kotthaus, D. Schuh, 
A. K. Rai, D. Reuter, A. D. Wieck and A. O. Govorov, Phys. Rev. B {\bf 83}, 165308 (2011).

\bibitem{2} W. Ketterle and N. J. van Druten, Phys. Rev. A {\bf 54}, 656 (1996).

\bibitem{3} D. S. Petrov, M. Holzmann, and G. V. Shlyapnikov, Phys. Rev. Lett.
{\bf 84}, 2551 (2000).

\bibitem{7} L. V. Butov, C. W. Lai, A. L. Ivanov, A. C. Gossard, and D. S.
Chemla, Nature (London) {bf 417}, 47 (2002).

\bibitem{4} D. P. Trauernicht, A. Mysyrowicz, and J. P. Wolfe, Phys. Rev. B {\bf 28},
3590 (1983).

\bibitem{5} K. Kash, J. M. Worlock, M. D. Sturge, P. Grabbe, J. P. Harbison,
A. Scherer, and P. S. D. Lin, Appl. Phys. Lett. {\bf 53}, 782 (1988).

\bibitem{6} V. Negoita, D. W. Snoke, and K. Eberl, Appl. Phys. Lett. {\bf 75}, 2059
(1999).

\bibitem{8} P. C. M. Christianen, F. Piazza, J. G. S. Lok, J. C. Maan, and
W. van der Vleuten, Physica B {\bf 249-251}, 624 (1998).

\bibitem{10} A. T. Hammack, M. Griswold, L. V. Butov, L. E. Smallwood, A. L.
Ivanov, and A. C. Gossard, Phys. Rev. Lett. {\bf 96}, 227402 (2006).

\bibitem{9} J. A. H. Stotz, R. Hey, P. V. Santos, and K. H. Ploog, Nature Mater.
{\bf 4}, 585 (2005).

\bibitem{11} S. Zimmermann, A. O. Govorov, W. Hansen, J. P. Kotthaus,
M. Bichler, and W. Wegscheider, Phys. Rev. B {\bf 56}, 13414 (1997).

\bibitem{12} S. Zimmermann, G. Schedelbeck, A. O. Govorov, A. Wixforth,
J. P. Kotthaus, M. Bichler, W. Wegscheider, and G. Abstreiter,
Appl. Phys. Lett. {\bf 73}, 154 (1998).

\bibitem{13} T. Huber, A. Zrenner, W.W egscheider, and M. Bichler, Phys. Status
Solidi A {\bf 166}, 5 (1998).

\bibitem{14} R. Rapaport, G. Chen, S. Simon, O. Mitrofanov, L. Pfeiffer, and
P. M. Platzman, Phys. Rev. B {\bf 72}, 075428 (2005).

\bibitem{15} V. B. Timofeev and A. V. Gorbunov, J. Appl. Phys. {\bf 101}, 081708
(2007).

\bibitem{16} A. T. Hammack, N. A. Gippius, S. Yang, G. O. Andreev, L. V.
Butov, M. Hanson, and A. C. Gossard, J. Appl. Phys. {\bf 99}, 066104
(2006).

\bibitem{17} G. Chen, R. Rapaport, L. N. Pffeifer, K. West, P. M. Platzman,
S. Simon, Z. V\"or\"os, and D. Snoke, Phys. Rev. B {\bf 74}, 045309
(2006).

\bibitem{18} A. G. Winbow, J. R. Leonard, M. Remeika, Y. Y. Kuznetsova,
A. A. High, A. T. Hammack, L. V. Butov, J. Wilkes, A. A. Guenther,
A. L. Ivanov, M. Hanson, and A. C. Gossard, Phys. Rev. Lett. {\bf 106}, 196806 (2011).

\bibitem{19} A. A. High, A. K. Thomas, G. Grosso,M.Remeika,A. T. Hammack,
A. D. Meyertholen, M. M. Fogler, L. V. Butov, M. Hanson, and
A. C. Gossard, Phys. Rev. Lett. {\bf 103}, 087403 (2009).

\bibitem{19b} H. Cruz, J. Appl. Phys. {\bf 113}, 153706 (2013).

\bibitem{20b} A. G. Winbow, J. R. Leonard, M. Remeika, Y. Y. Kuznetsova, A. A. High,
A. T. Hammack, L. V. Butov, J. Wilkes, A. A. Guenther, A. L. Ivanov, M. Hanson and
A. C. Gossard, Phys. Rev. Lett. {\bf 106}, 196806 (2011).

\bibitem{29} D. Luis, H. Cruz and N. E. Capuj, Phys. Rev. B {\bf 59}, 9787
(1999); D. Luis, J. P. D\'\i az, N. E. Capuj and H. Cruz, J. Appl. Phys. 
{\bf 88}, 943 (2000).

\bibitem{30} H. Cruz, J. Appl. Phys. {\bf 83,} 2677 (1998); J. Phys.:
Condens. Matter {\bf 10,} 4677 (1998); Phys. Rev. B {\bf 65}, 245313 (2002).

\bibitem{27} H. G. Roskos, M. C. Nuss, J. Shah, K. Leo, D. A. B. Miller, A.
M. Fox, S. Schmitt-Rink, and K. K\"ohler, Phys. Rev. Lett. {\bf 68}, 2216
(1992).

\bibitem{32} H. Cruz, D. Luis, N. E. Capuj and L. Pavesi, J. Appl. Phys. {\bf %
83}, 7693 (1998); D. Luis, J. P. D\'\i az and H. Cruz, Phys. Rev. B {\bf 62}%
, 7264 (2000).

\bibitem{33} H. De Raedt, Comp. Phys. Rep. {\bf 7}, 1 (1987); M. O. Vassel
and J. Lee, Phys. Rev. B {\bf 44}, 3864 (1991).
\end{thebibliography}
\end{document}